\documentclass[11pt]{article}  %diffuse/cospar2000/paper
\usepackage{fleqn,cospar}
\usepackage{url}
\usepackage{graphicx}

\usepackage[figuresright]{rotating}

\onecolumn

\newcommand{\pub}[4]{{\it #1} {\bf #2}, #3, #4.}
\newcommand\subtitle{
\vspace{-57mm} \vspace{-1\baselineskip}
\noindent \mbox{\scriptsize \it
Invited talk at the 33rd COSPAR Scientific Assembly (2000). To appear in Adv.\ Space Res.\ 2001
}\vspace{55mm}}

\title{\bf MODELS FOR GALACTIC COSMIC-RAY PROPAGATION}

\author{A.W.~Strong\address{Max-Planck-Institut f\"ur extraterrestrische Physik, 
Postfach 1312, 85741 Garching, Germany}
   and
I.V.~Moskalenko\address{NASA/Goddard Space Flight Center, Code 660, %
Greenbelt, MD 20771, USA}%
\thanks{NRC Senior Research Associate}%
\thanks{on leave from Institute of Nuclear Physics, %
M.V.Lomonosov Moscow State University, 
Moscow 119899, Russia}}

\begin{document}

% typeset front matter
\maketitle
\subtitle

\begin{abstract}
A new numerical model of particle propagation in the Galaxy
has been developed, which allows the study of cosmic-ray and gamma-ray production 
and propagation in 2D or 3D, including a full reaction network. 
This is a further development of
the code which has been used for studies of cosmic ray reacceleration,
Galactic halo size, antiprotons and positrons in cosmic rays,
the interpretation of diffuse continuum gamma rays, and dark matter.
In this paper we illustrate recent results focussing on B/C, sub-Fe/Fe,
ACE radioactive isotope data, source abundances and antiprotons.  
From the radioactive nuclei we derive a range of 3--7 kpc for the
height of the cosmic-ray halo.

%{\em Advances in Space Research}
\end{abstract}

\section*{INTRODUCTION}
In recent years, new and accurate data have become available in 
cosmic-ray (CR) astrophysics,
and much progress in understanding the origin and propagation 
of CR can be made, but this requires a systematic and 
self-consistent approach. 
With this motivation we have developed a 
numerical propagation code which aims to reproduce 
observational data of many kinds related to CR, such 
as direct measurements of nuclei, antiprotons, positrons and electrons, gamma 
rays and synchrotron radiation.

In the first version, the model was cylindrically symmetric and was written 
in fortran-90. This version has proven that the original goal, a ``realistic 
standard model'' of CR production and propagation, is possible to achieve. 
It has been used for 
studies of CR reacceleration, halo size, production and propagation of 
positrons and antiprotons, dark matter annihilation, and the interpretation 
of diffuse continuum gamma rays. Many aspects, however, cannot be addressed 
in such a model, e.g.\ the stochastic nature of the CR sources in space and 
time which is important for high-energy electrons with short cooling times, 
and local inhomogeneities in the gas density which can affect radioactive 
secondary  ratios in CR.

The experience gained from the original version allowed us to design a new 
version of the model, entirely rewritten in C++, which is much more flexible, 
which incorporates essential improvements over  the older model, and 
in which a  3-dimensional spatial grid can be employed. In addition it is now 
possible to solve the full nuclear reaction network on the spatially resolved 
grid.  We keep however an option which allows us to switch 
to the older cylindrically symmetrical (``2D'') model  since this is
still a sufficient approximation for nuclei in 
the absence of spatial fluctuations and
is much faster to compute than the full 3D case.

The code can thus serve as a
complete substitute for the conventional ``leaky-box'' or ``weighted-slab''
propagation models usually employed, with many associated advantages
such as the correct treatment of radioactive nuclei, realistic gas and
source distributions etc.  The present work 
illustrates the use of the new 2D code to interpret
B/C, sub-Fe/Fe,  new radioactive isotopic 
ratio data from ACE, source abundances and antiprotons.

\section*{MODELS}

The GALPROP models have been described in full 
detail elsewhere (Strong and Moskalenko 1998); 
here we just summarize briefly their basic features.
The results obtained with the original version of GALPROP have been discussed
in a recent review (Strong and Moskalenko 1999).

The 2D models have cylindrical symmetry in the
Galaxy, and the basic coordinates are $(R,z,p)$ where $R$ is
Galactocentric radius, $z$ is the distance from the Galactic plane and
$p$ is the total particle momentum. The  propagation
region is bounded by $R=R_h$, $z=\pm z_h$ beyond which free escape is
assumed. We take $R_h=30$ kpc. For a given $z_h$ the diffusion
coefficient as a function of momentum  and the reacceleration
parameters is determined by the energy-dependence of the B/C ratio.  
Reacceleration provides a natural
mechanism to reproduce the B/C ratio without an ad-hoc form for the
diffusion coefficient.  The spatial diffusion coefficient is taken as
$\beta D_0(\rho/\rho_0)^\delta$, assuming independence of position, 
where $\rho$ is rigidity. 
We use values of $\delta$ near but not necessarily equal to the 
Kolmogorov value of $\delta=1/3$. For the case of
reacceleration the momentum-space diffusion coefficient $D_{pp}$ is
related to the spatial coefficient (e.g.\ Seo and Ptuskin 1994).
The reacceleration is parameterized by $v_A^2/w$
where $v_A$ is the Alfv\'en speed and $w$ the ratio of wave energy 
density to magnetic field energy density (Seo and Ptuskin 1994).
The injection spectrum of nuclei is assumed
to be a power law in momentum, $dq(p)/dp \propto p^{-\gamma}$ for the
injected particle density, if necessary with a break.

The interstellar hydrogen distribution uses HI and CO surveys and
information on the ionized component; the Helium fraction of the gas is
taken as 0.11 by number.  Energy losses of
nuclei by ionization and Coulomb interactions
are included. 

The distribution of CR
sources is chosen to reproduce the CR distribution
determined by analysis of EGRET gamma-ray data (Strong and Mattox 1996). The
source distribution adopted was described in Strong and Moskalenko (1998).  It
adequately reproduces the observed gamma-ray based gradient, while being
significantly flatter than the observed distribution of supernova
remnants.

%%% Figure 1 %%%

\begin{figure}[tb]
\centerline{\includegraphics[width=120mm]{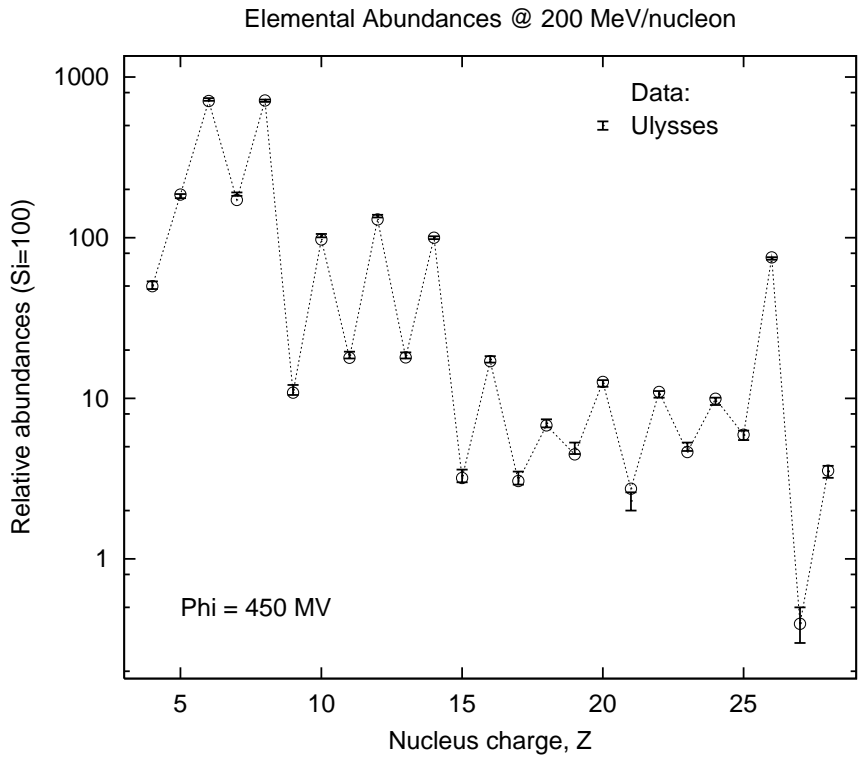}} \vspace{-1em}
\caption{Relative CR propagated elemental abundances as calculated in our model at
200 MeV/nucleon in the heliosphere -- circles. Modulation potential $\Phi=450$ MV. 
Data: Duvernois and Thayer (1996) -- vertical bars.}
\end{figure}

The primary source abundances are adjusted to give as good agreement as possible 
with the observed abundances after propagation (Figure 1), 
for a given set of cross-sections.
The heliospheric modulation is taken into account using the force-field 
approximation.

In the new version, apart from the option of a full 3D treatment, we have
updated the cross-section code to include latest measurements
and energy dependent fitting functions. The nuclear reaction network
is built using the Nuclear Data Sheets. The isotopic cross section database 
consists of more than 2000 points collected from sources published in 1969--1999.
This includes a critical re-evaluation of some data and cross checks.
The isotopic cross sections for B/C were calculated using the authors' fits to major 
beryllium and boron production cross sections C,N,O $\to$ Be,B.
Other cross sections are calculated using 
the Webber et al.\ (1990)\footnote{
Code \url{WNEWTR.FOR} of 1993 as  posted at
\url{http://spdsch.phys.lsu.edu/SPDSCH_Pages/
Software_Pages/Cross_Section_Calcs/WebberKishSchrier.html}}
and/or Silberberg and Tsao\footnote{
Code \url{YIELDX_011000.FOR} of 2000 as  posted at
\url{http://spdsch.phys.lsu.edu/SPDSCH_Pages/Software_Pages/Cross_Section_Calcs/
SilberburgTsao.html}}
phenomenological approximations renormalized to the data where it exists.
In this paper we use our fits combined with the Webber et al.\ approximation
renormalized to available data. 

The reaction network is solved starting at the heaviest nuclei (i.e.\ $^{64}$Ni),
solving the propagation equation, computing all the resulting secondary source
functions, and proceeding to the nuclei with $A-1$. 
The procedure is repeated down to $A=1$.
In this way all secondary, tertiary etc.\ reactions are 
automatically accounted for.
To be completely accurate for all isotopes, e.g.\ for some rare 
cases of $\beta^\pm$-decay, the whole loop is repeated twice.

\section*{PROPAGATION OF COSMIC RAYS}

From the point of view of isotope ratios, the main advance here is our employment 
of the full reaction network.
For this purpose the 2D model is sufficient.
 
Figure 2 shows the predicted local interstellar (LIS)
and modulated B/C ratio compared with observations;
the reacceleration reproduces the peak quite well.
The new ACE data (Davis et al.\ 2000) confirm
the presence of the peak and give an improved basis for fitting the 
reacceleration. 
Note that the modulation potential estimated for the ACE data by 
Davis et al.\ is $\Phi=350$ MV; 
our calculations made with $\Phi=400 - 500$ MV agree with the 
ACE data, and we adopt 450 MV in what follows.
The model we consider has a
diffusion coefficient index $\delta=0.36$ and injection spectrum
index $\gamma=2.35$.
We find that a value of $v_A/ w^{1/2} =31$ km s$^{-1}$ gives the best match. 
A ``Kolmogorov'' model with $\delta=1/3$ ($v_A/w^{1/2} =24$ km s$^{-1}$)
gives less satisfactory agreement with the high-energy data.

Figure 3 shows 
(Sc+Ti+V)/Fe for the same model. The agreement is quite good given that 
the model was not fitted to this ratio; this ratio favours $\delta=0.36$,
and less satisfactorily matched by $\delta=1/3$.
%Note that the good match (also in Figure 1) is obtained by adding to
%the source abundance of Ti which appears to be a factor of ten higher
%than the solar one (see Figure 9 and discussion below).
%In this sense the pure secondary to primary ratio (Sc+V)/Fe is less
%adjustable and would therefore be preferable for comparisons. 

%%% Figure 2 %%%
%%% Figure 3 %%%

\begin{figure*}[!tb]
\begin{minipage}[t]{88mm}
\includegraphics[width=87mm]{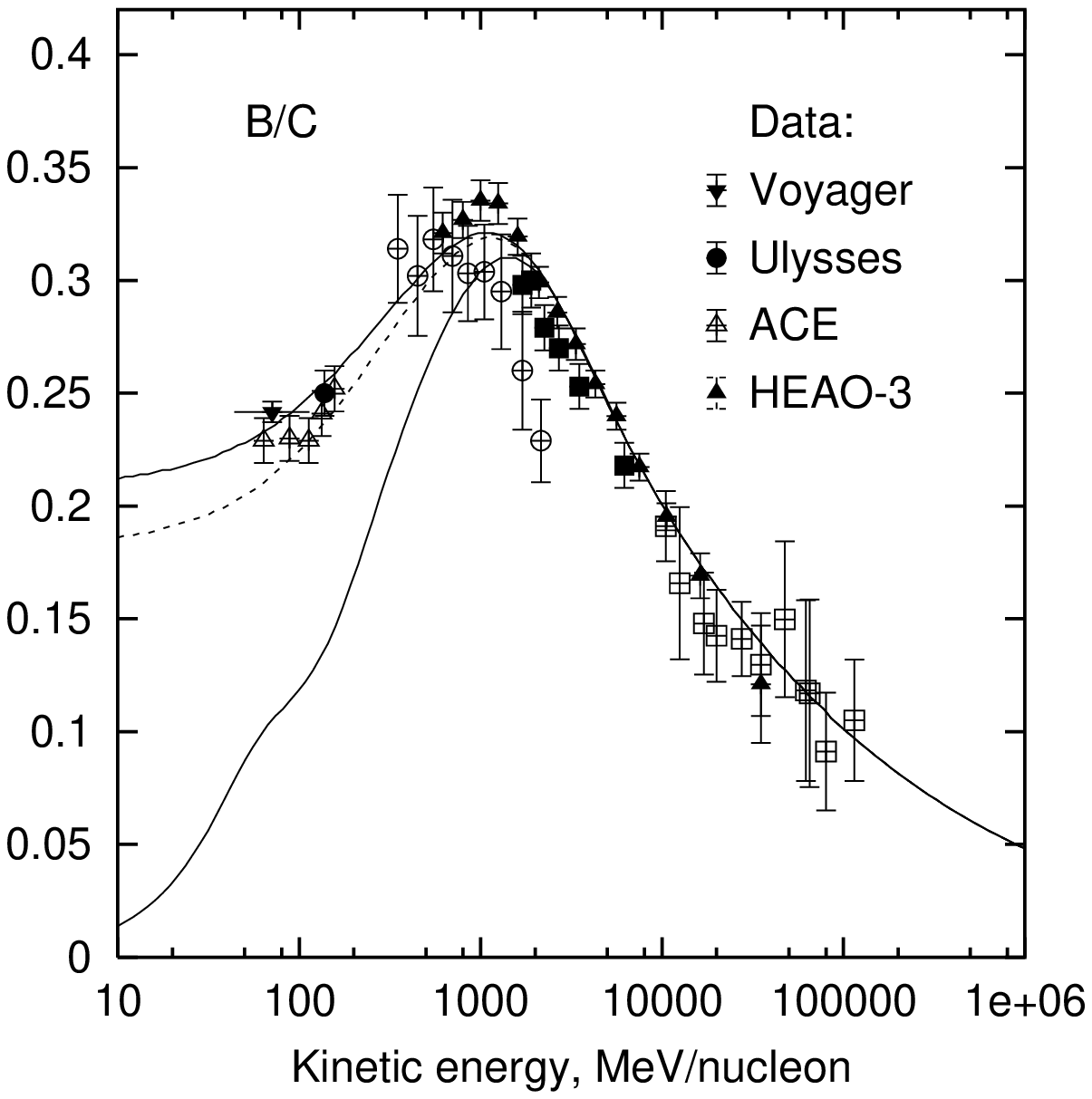}
\vspace{-0.5\baselineskip}
\caption{B/C ratio  calculated  for $z_h=4$ kpc. 
Lower curve LIS, upper modulated. Model with
$(\delta,\gamma)=(0.36,2.35)$: solid curve -- $\Phi=500$ MV,
dotted curve -- $400$ MV.
Data below 200 MeV/nucleon: ACE -- Davis et al.\ (2000),
Ulysses -- DuVernois et al.\ (1996), Voyager -- Lukasiak et al.\ (1999);
high energy data: HEAO-3 -- Engelmann et al.\ (1990), 
for other references see Stephens and Streitmatter (1998).}
\end{minipage}
\hspace{\fill}
\begin{minipage}[t]{88mm}
\includegraphics[width=87mm]{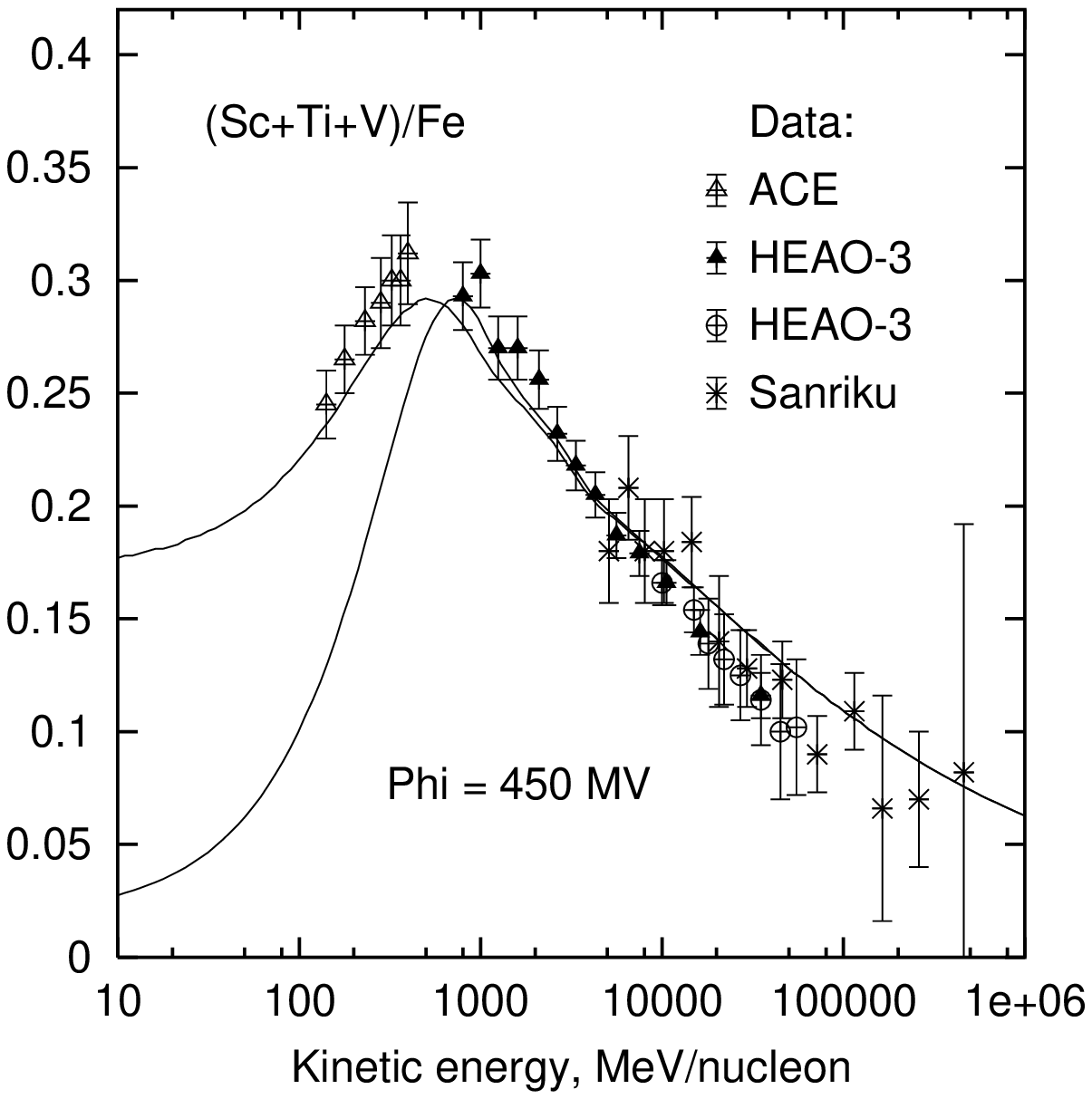}
\vspace{-0.5\baselineskip}
\caption{SubFe/Fe ratio  calculated  for $z_h=4$ kpc. 
Lower curve LIS, upper modulated. Model with
$(\delta,\gamma)=(0.36,2.35)$. Data: ACE -- Davis et al.\ (2000),
HEAO-3 -- Binns et al.\ (1988) and Engelmann et al.\ (1990),
Sanriku experiment -- Hareyama et al.\ (1999). We used  HEAO-3 data only below 
45 GeV because of absence of vanadium measurements above this energy.}
\end{minipage}
\end{figure*}

%%% Figure 4 %%%
%%% Figure 5 %%%
%%% Figure 6 %%%
%%% Figure 7 %%%

\begin{figure*}[!tb]
\begin{minipage}[t]{88mm}
\includegraphics[width=87mm]{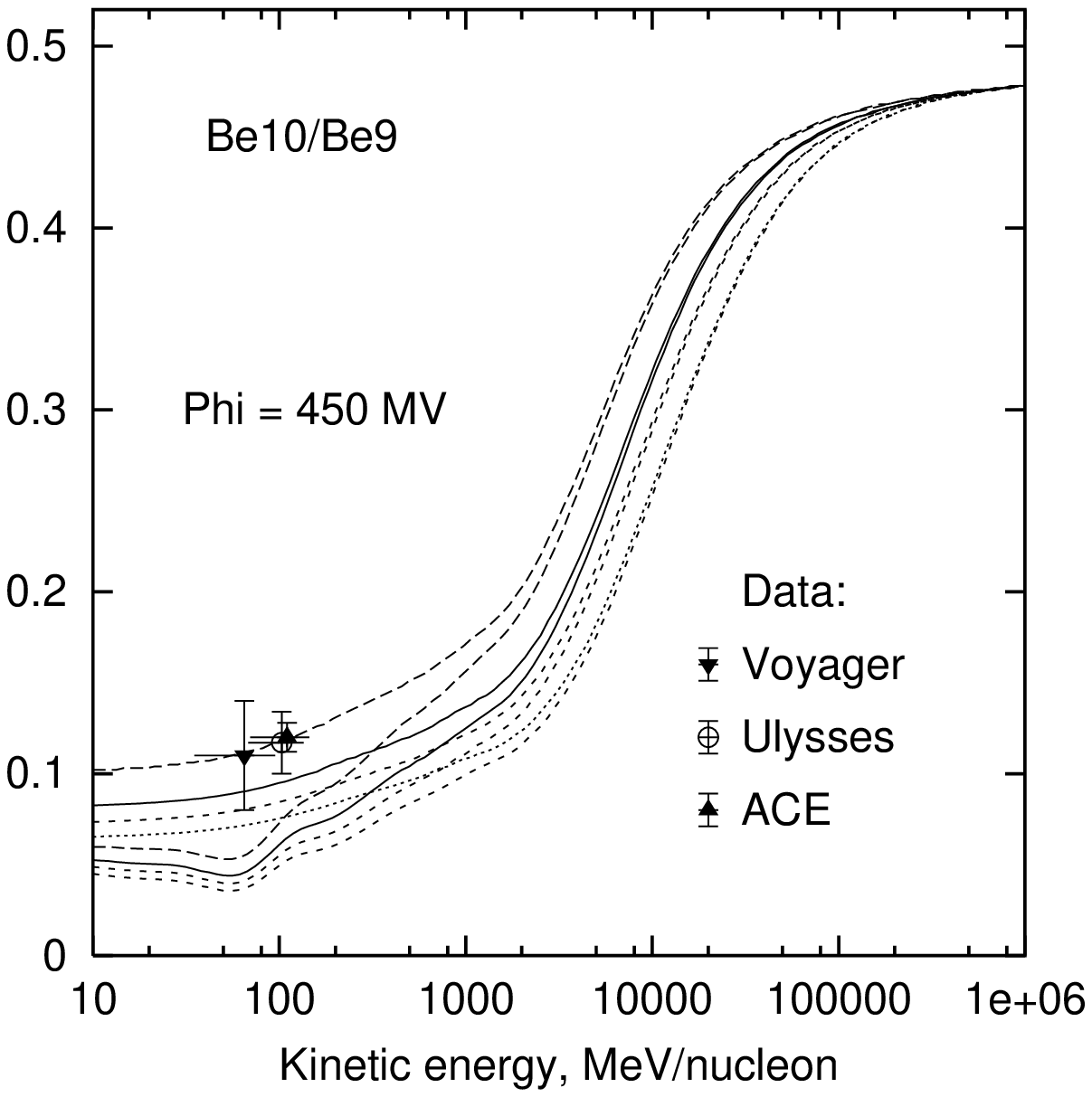} \vspace{-1em}
\caption{$^{10}$Be/$^9$Be ratio  calculated  for $z_h=2, 4, 6, 10$ kpc
(top to bottom). Lower curves LIS, upper modulated. Data:
ACE -- Yanasak et al.\ (2000), Ulysses -- Connell (1998), 
Voyager -- Lukasiak et al.\ (1999).}
\end{minipage}
\hspace{\fill}
\begin{minipage}[t]{88mm}
\includegraphics[width=87mm]{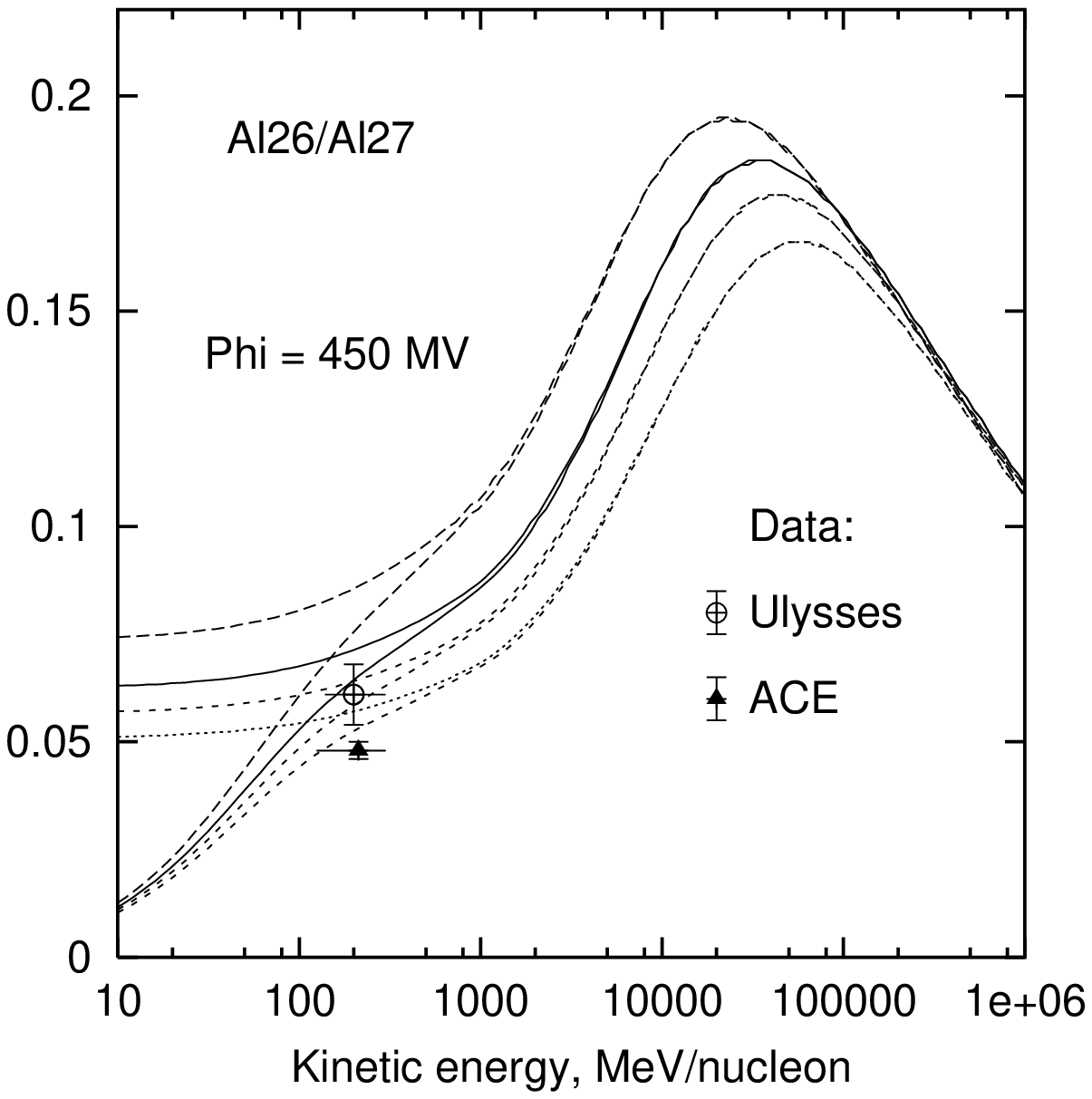} \vspace{-1em}
\caption{$^{26}$Al/$^{27}$Al ratio  calculated  for $z_h=2, 4, 6, 10$ kpc
(top to bottom). Lower curves LIS, upper modulated. Data:
ACE -- Yanasak et al.\ (2000), Ulysses -- Simpson and Connell (1998).
}
\end{minipage}
%\end{figure*}
%
\vskip 1cm
%\begin{figure*}[!tb]
\begin{minipage}[t]{88mm}
\includegraphics[width=87mm]{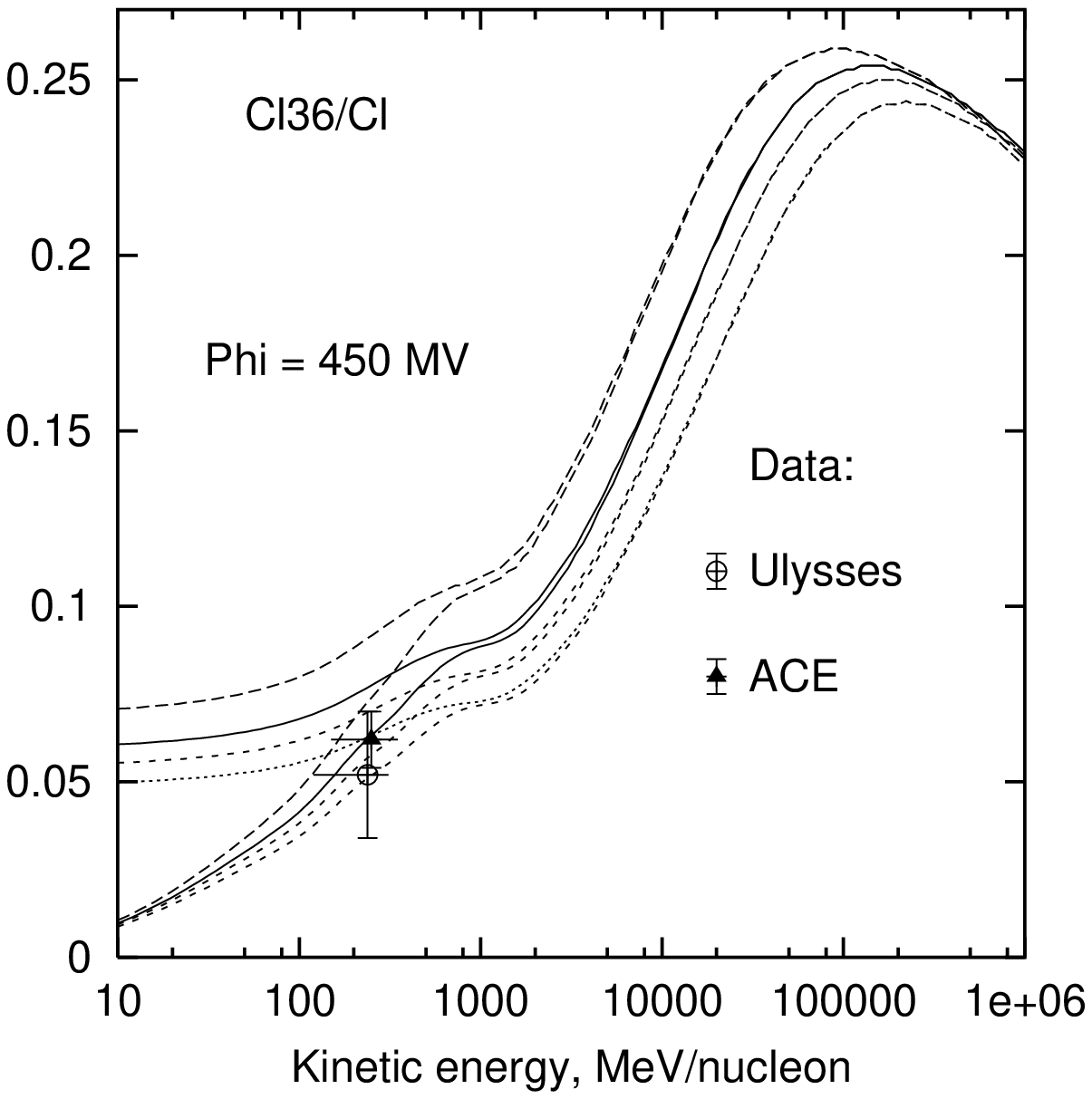} \vspace{-1em}
\caption{$^{36}$Cl/Cl ratio  calculated  for $z_h=2, 4, 6, 10$ kpc
(top to bottom). Lower curves LIS, upper modulated. Data:
ACE -- Yanasak et al.\ (2000), Ulysses -- Connell et al.\ (1998).}
\end{minipage}
\hspace{\fill}
\begin{minipage}[t]{88mm}
\includegraphics[width=87mm]{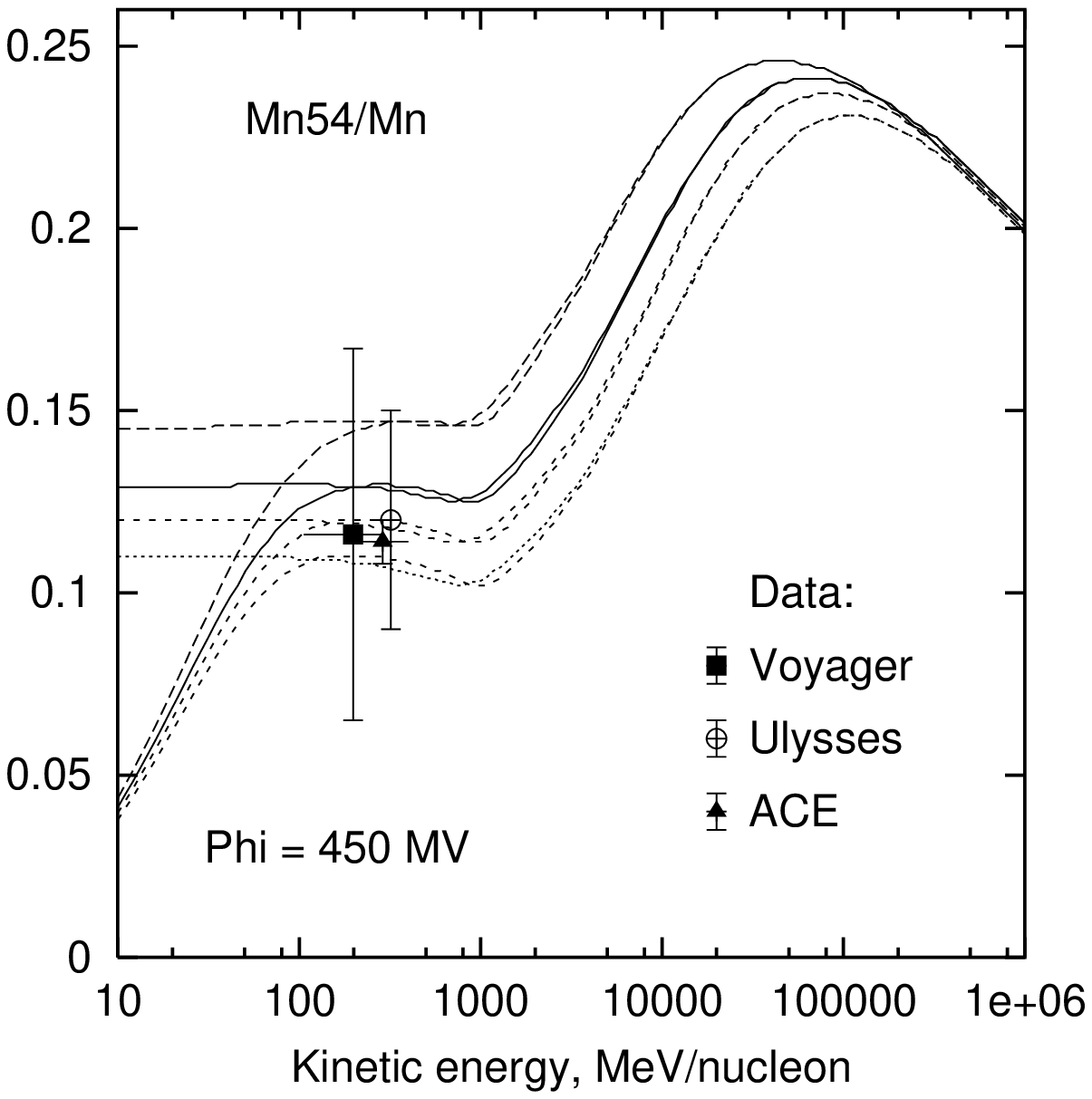} \vspace{-1em}
\caption{$^{54}$Mn/Mn ratio  calculated  for $z_h=2, 4, 6, 10$ kpc
(top to bottom). Lower curves LIS, upper modulated. Data:
ACE -- Yanasak et al.\ (2000), Ulysses -- DuVernois (1997), 
Voyager -- Lukasiak et al.\ (1997).}
\end{minipage}
%\vspace{-1.5\baselineskip}
\end{figure*}

As in previous work we study the halo size $z_h$ using $^{10}$Be and 
other radioactive isotopes. For each $z_h$ the model is adjusted to fit B/C
and then the radioactive nuclei  are compared with data.
Figure 4 shows $^{10}$Be/$^{9}$Be;
notice that the rapid falloff in the ratio at low energies, 
which is caused by the variation in the $^{12}$C$\to^{10}$Be cross-section, 
is critical, and makes this ratio rather sensitive to the modulation. 
Derivation of $z_h$ from $^{10}$Be/$^{9}$Be is discussed below.
Observations at higher energies, where modulation is smaller, 
will greatly help in this regard.

For the heavier radioactive nuclei the procedure is first to adjust the
source abundances at the appropriate energy to agree with the relevant
observed stable nuclei ratios and then derive the fluxes of the
radioactive isotopes. In this way the uncertainty in the denominator of
the ratio is reduced.  Figures 5--7 show
$^{26}$Al/$^{27}$Al, $^{36}$Cl/Cl and $^{54}$Mn/Mn.
The ACE data points imply  halo sizes of a few kpc.
Some uncertainty arises from the error bars on the elemental abundance
measurements from Ulysses, to which we tune our propagated abundances.
The dominant uncertainty comes however from the cross-sections and the
modulation, while the experimental values for the ratios measured by ACE
are rather accurate. 

We now try to take into account the cross section and abundance errors. 
The calculated abundances of the four radioactive isotopes at \linebreak[3] 200 MeV/nucleon
(100 MeV/nucleon for Be) and $\Phi=450$ MV are shown in Table 1 for different halo sizes. 
The ratios in the Table are calculated using Ulysses elemental abundances 
(DuVernois and Thayer 1996) normalized to silicon (Si=100): 
Al $=18.5\pm0.8$, Cl $=3.2\pm0.3$, Mn $=5.9\pm0.4$. 
This removes some uncertainty in the denominator for these ratios.
The first error bar was obtained from the elemental
abundance error, the second error bar is larger and obtained 
assuming 20\% error in the production
cross sections of $^{36}$Cl and $^{54}$Mn, and 35\% 
error for $^{26}$Al. When deriving the
halo size these  translate into a wide range of halo sizes
since the difference between calculated abundances for
4 and 6 kpc, and 6 and 10 kpc halos is only about 10\%.
Assuming the errors in Table 1 are independent, the total error estimate may thus be  30\% or even larger, making the halo size estimates 
all consistent with a lower limit of $\sim3$ kpc.

The case of Be is slightly different. All Be isotopes are purely secondary,
so that we cannot tune their abundances by varying the source abundance. 
While the elemental abundance 
agrees well with Ulysses measurements, the relative isotopic fraction of $^7$Be 
is 10\% less and $^9$Be is 15\% more than measured (Connell 1998):
$^7$Be = 0.561, $^9$Be = 0.393, $^{10}$Be = 0.046.
The Be isotope production cross sections, we believe, are more 
accurate than those for  most  other nuclei
since we use fits to the major production channels.
We therefore take the largest deviation (15\%) as an estimate
of $^{10}$Be production cross section error. The ratio in Table 1
is obtained using the measured fraction of $^9$Be.
The corrected value then gives the range  $z_h = 1.5 - 6.0$ kpc.

\begin{table*}[!t]%=====================================================+
%\vspace{-1\baselineskip}
\caption{Radioactive isotopic abundances (Si=100)$^*$.}
\small %\footnotesize
\begin{tabular}{rcccccccc}
\hline \hline \noalign{\smallskip}
$z_h$  & $^{10}$Be& $^{10}$Be/$^9$Be$^\dagger$, $10^{-1}$
       & $^{26}$Al& $^{26}$Al/$^{27}$Al, $10^{-2}$
       & $^{36}$Cl& $^{36}$Cl/Cl, $10^{-2}$
       & $^{54}$Mn& $^{54}$Mn/Mn, $10^{-1}$ \\
\hline \noalign{\smallskip}
 2 kpc & 2.27     & 1.33 $_{-0.17}^{+0.20}$
       & 1.44     & 8.44 $_{-0.38}^{+0.42}$ $_{-2.32}^{+3.30}$  %$_{-1.50}^{+1.86}$ #20% err#
       & 0.268    & 8.38 $_{-0.72}^{+0.87}$ $_{-1.40}^{+1.68}$
       & 0.900    & 1.53 $_{-0.10}^{+0.11}$ $_{-0.25}^{+0.31}$ \\
\noalign{\smallskip}
 4 kpc & 1.84     & 1.09 $_{-0.14}^{+0.16}$
       & 1.20     & 6.91 $_{-0.33}^{+0.31}$ $_{-1.88}^{+2.65}$  %$_{-1.22}^{+1.50}$
       & 0.225    & 7.04 $_{-0.60}^{+0.73}$ $_{-1.17}^{+1.41}$
       & 0.769    & 1.30 $_{-0.08}^{+0.10}$ $_{-0.22}^{+0.26}$ \\
\noalign{\smallskip}
 6 kpc & 1.62     & 0.97 $_{-0.13}^{+0.15}$
       & 1.07     & 6.16 $_{-0.27}^{+0.30}$ $_{-1.67}^{+2.34}$  %$_{-1.08}^{+1.32}$
       & 0.203    & 6.35 $_{-0.54}^{+0.66}$ $_{-1.06}^{+1.27}$
       & 0.697    & 1.18 $_{-0.08}^{+0.09}$ $_{-0.20}^{+0.24}$ \\
\noalign{\smallskip}
10 kpc & 1.41     & 0.87 $_{-0.11}^{+0.13}$
       & 0.952    & 5.42 $_{-0.24}^{+0.26}$ $_{-1.46}^{+2.04}$  %$_{-0.94}^{+1.16}$
       & 0.179    & 5.60 $_{-0.48}^{+0.58}$ $_{-0.93}^{+1.12}$
       & 0.624    & 1.06 $_{-0.07}^{+0.08}$ $_{-0.18}^{+0.21}$ \\
\noalign{\smallskip}
\multicolumn{2}{l}{ACE data$^\ddagger$}
                  & 1.20 $\pm0.08$
      &           & 4.8  $\pm0.2$
      &           & 6.2  $\pm0.8$
      &           & 1.14 $\pm0.06$ \\
\noalign{\smallskip} \hline 
\end{tabular} \smallskip \\
% Ulysses: Al=18.5+/-0.8; Cl=3.2+/-0.3; Mn=5.9+/-0.4
{\footnotesize 
$^*$The error bars derived (i) using the elemental abundance
error, (ii) assuming 20\% error in the production cross sections of $^{36}$Cl 
and $^{54}$Mn, and 35\% error for $^{26}$Al.\\
$^\dagger$Error bars derived assuming 15\% error in $^{10}$Be production cross section.\\
$^\ddagger$Yanasak et al.\ (2000).}
\vspace{-1.5\baselineskip}
\end{table*}%=====================================================+

%%% Figure 8 %%%
%%% Figure 9 %%%

\begin{figure*}[!t]
\begin{minipage}[t]{88mm}
\includegraphics[width=87mm,height=70mm]{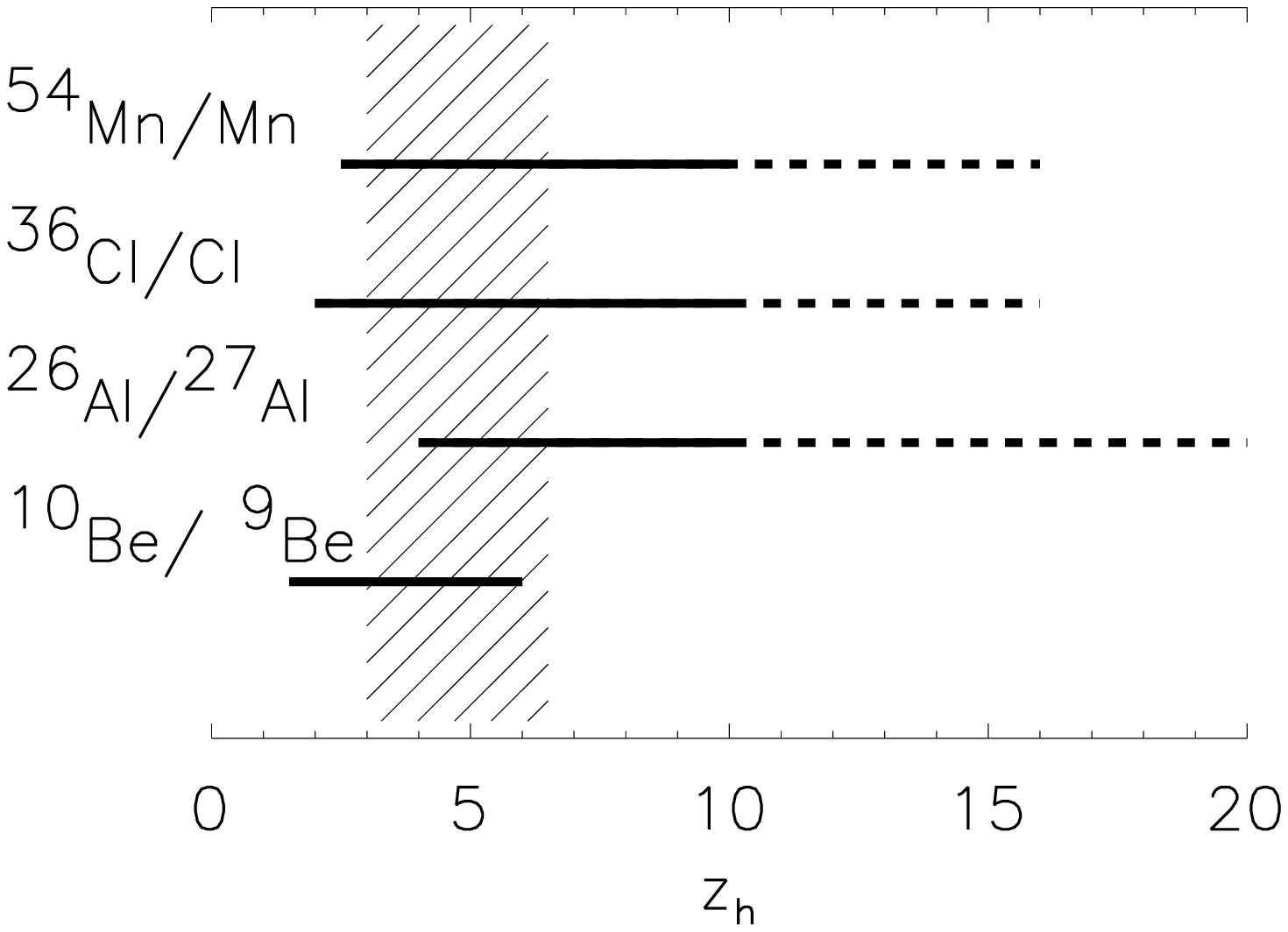} \vspace{-1em}
\caption{Halo size limits as derived from the abundances of the four 
radioactive isotopes (Table 1) and ACE data. 
The ranges reflect errors in ratio measurements, source abundances 
and production cross-sections. Dashed lines indicate uncertain upper limits.
The shaded area indicates the range consistent with all ratios.}
\end{minipage}
\hspace{\fill}
\begin{minipage}[t]{88mm}
\centerline{\includegraphics[width=70mm]{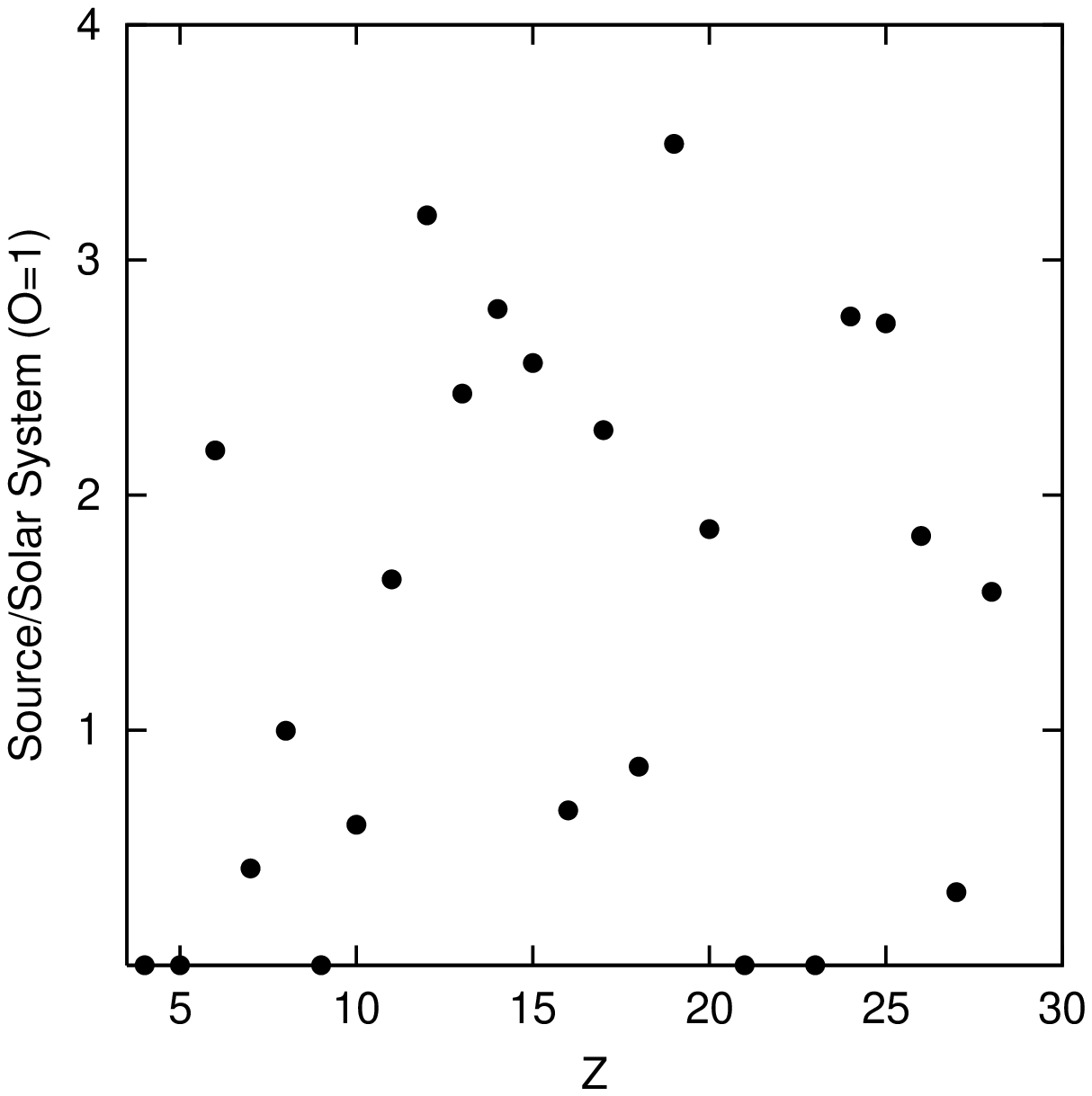}} \vspace{-1em}
\caption{Elemental source/solar system abundance ratio, ([O] = 1).
Solar system abundances are from Anders and Grevesse (1989).}
\end{minipage}
\end{figure*}

Figure 8 summarizes the halo size constraints based on the four
radioactive isotopes considered using ACE data (Yanasak et al. 2000).
These have been calculated  by requiring  consistency
of the calculated ratio  with the ACE data
taking into account the error bars on both prediction and data.
The conclusion to be drawn from
all radioactive nuclei is that we require $z_h=3-7$ kpc,
i.e.\ a large halo, at least within the context of the models
considered here.  This is consistent with our previous result $z_h=4-10$ kpc 
(Strong and Moskalenko 1998) based on Ulysses data (Connell 1998), 
and now the ACE data lead to a more robust estmate. 
It is also consistent with other results, e.g., 
Webber and Soutoul (1998) found $z_h=2 - 4$ kpc from HEAO-3,
and Ptuskin and Soutoul (1998) give $4.9_{-2}^{+4}$ kpc.

\begin{table*}[!t]%=====================================================+
%\vspace{-1\baselineskip}
\caption{Propagated and source isotopic abundances ($^{16}$O=100).}
\footnotesize
\begin{minipage}[t]{48mm}
\begin{tabular}{llcc|}
\hline \hline \noalign{\smallskip}
   &      & Prop.   &  Source    \\ % &Source/SS\\
   & Z, A & abund.  &  abund.     \\ % & abundances \\
\hline \noalign{\smallskip}
Be &\ 4, 7 & $  3.62    $ & $   -      $\\%&     -      \\
   &\ 4, 9 & $  3.21    $ & $   -      $\\%&     -      \\
   &\ 4,10 & $ 0.337    $ & $   -      $\\%&     -      \\
B  &\ 5,10 & $  7.46    $ & $   -      $\\%&     -      \\
   &\ 5,11 & $  19.2    $ & $   -      $\\%&     -      \\
C  &\ 6,12 & $  94.2    $ & $  69.4    $\\%&    2.19    \\
   &\ 6,13 & $  7.66    $ & $ 0.681    $\\%&    1.77    \\
N  &\ 7,14 & $  12.0    $ & $  4.76    $\\%&   0.414    \\
   &\ 7,15 & $  12.6    $ & $   -      $\\%&     -      \\
O  &\ 8,16 & $  100.    $ & $  100.    $\\%&    1.00    \\
   &\ 8,17 & $  1.21    $ & $   -      $\\%&     -      \\
   &\ 8,18 & $  1.37    $ & $   -      $\\%&     -      \\
F  &\ 9,19 & $  1.55    $ & $   -      $\\%&     -      \\
Ne &10,20 & $  8.24    $ & $  8.44    $\\%&   0.523    \\
   &10,21 & $  1.52    $ & $   -      $\\%&     -      \\
   &10,22 & $  4.21    $ & $  2.53    $\\%&    1.17    \\
Na &11,23 & $  2.56    $ & $ 0.572    $\\%&    1.64    \\
Mg &12,24 & $  13.3    $ & $  17.0    $\\%&    3.19    \\
   &12,25 & $  2.66    $ & $  2.20    $\\%&    3.06    \\
   &12,26 & $  2.67    $ & $  2.74    $\\%&    3.33    \\
Al &13,26 & $ 0.172    $ & $   -      $\\%&     -      \\
   &13,27 & $  2.40    $ & $  1.47    $\\%&    2.43    \\
Si &14,28 & $  12.1    $ & $  18.5    $\\%&    2.72    \\
   &14,29 & $  1.48    $ & $  1.63    $\\%&    4.54    \\
   &14,30 & $ 0.727    $ & $ 0.545    $\\%&    2.23    \\
P  &15,31 & $ 0.458    $ & $ 0.218    $\\%&    2.56    \\
\noalign{\smallskip} \hline
\end{tabular}
\end{minipage}
\begin{minipage}[t]{65.9mm}
\begin{tabular}{llcc|}
\hline \hline \noalign{\smallskip}
   &      & Propagated  &  Source    \\ % &Source/SS\\
   & Z, A & abundances  & abundances \\ % & abundances \\
\hline \noalign{\smallskip}
S  &16,32 & $  1.73    $ & $  2.64    $\\%&   0.640    \\
   &16,33$^*$
          & $ 0.305    $ & $ 5.717\times10^{-2}$\\%&    1.68    \\
   &16,34 & $ 0.387    $ & $ 0.172    $\\%&   0.882    \\
   &16,36 & $ 2.251\times10^{-2}$ & $   -      $\\%&     -      \\
Cl$^*\!$
   &17,35 & $ 0.252    $ & $ 2.450\times10^{-2}$\\%&   0.670    \\
   &17,36 & $ 3.289\times10^{-2}$ & $   -      $\\%&     -      \\
   &17,37 & $ 0.153    $ & $ 8.712\times10^{-2}$\\%&    6.98    \\
Ar &18,36 & $ 0.399    $ & $ 0.681    $\\%&   0.844    \\
   &18,37 & $ 0.148    $ & $   -      $\\%&     -      \\
   &18,38 & $ 0.364    $ & $ 0.136    $\\%&   0.849    \\
   &18,40 & $ 6.236\times10^{-2}$ & $   -      $\\%&     -      \\
K  &19,39 & $ 0.321    $ & $ 0.136    $\\%&    3.77    \\
   &19,40 & $ 0.143    $ & $   -      $\\%&     -      \\
   &19,41 & $ 0.178    $ & $   -      $\\%&     -      \\
Ca &20,40 & $ 0.578    $ & $  1.17    $\\%&    1.87    \\
   &20,41$^*$
          & $ 0.126    $ & $ 2.723\times10^{-2}$\\%&     -      \\
   &20,42 & $ 0.357    $ & $   -      $\\%&     -      \\
   &20,43 & $ 0.374    $ & $   -      $\\%&     -      \\
   &20,44 & $ 0.375    $ & $   -      $\\%&     -      \\
   &20,46 & $ 3.037\times10^{-3}$ & $   -      $\\%&     -      \\
%   &20,48 & $   -      $ & $   -      $\\%&     -      \\
Sc &21,45 & $ 0.392    $ & $   -      $\\%&     -      \\
Ti$^*\!$
   &22,44 & $ 1.264\times10^{-2}$ & $   -      $\\%&     -      \\
   &22,46 & $ 0.375    $ & $   -      $\\%&     -      \\
   &22,47 & $ 0.478    $ & $ 6.806\times10^{-2}$\\%&     -      \\
   &22,48 & $ 0.588    $ & $ 0.272    $\\%&    3.65    \\
   &22,49$^\ddagger$
          & $ 7.615\times10^{-2}$ & $   -      $\\%&     -      \\
\noalign{\smallskip} \hline
\end{tabular}
\end{minipage}
\begin{minipage}[t]{66mm}
\begin{tabular}{llcc}
\hline \hline \noalign{\vspace{1pt}}
   &      & Propagated  &  Source     \\%&Source/SS\\
   & Z, A & abundances  & abundances  \\%& abundances \\
\noalign{\vspace{2pt}}
\hline \noalign{\vspace{2pt}}
Ti$^*\!$
   &22,50 & $ 3.853\times10^{-2}$ & $ 4.901\times10^{-2}$\\%&    3.23    \\
V  &23,49$^\dagger$
          & $ 0.368    $ & $   -      $\\%&     -      \\
   &23,50 & $ 0.224    $ & $   -      $\\%&     -      \\
   &23,51$^\ddagger$
          & $ 7.123\times10^{-2}$ & $   -      $\\%&     -      \\
Cr &24,50$^*$ 
          & $ 0.235    $ & $ 5.990\times10^{-2}$\\%&    7.81    \\
   &24,51$^\dagger$
          & $ 0.377    $ & $   -      $\\%&     -      \\
   &24,52 & $ 0.609    $ & $ 0.286    $\\%&    1.85    \\
   &24,53 & $ 0.160    $ & $ 0.127    $\\%&    7.05    \\
   &24,54$^{*\ddagger}\!$
          & $ 4.416\times10^{-2}$ & $ 3.812\times10^{-2}$\\%&    8.34    \\
Mn &25,53 & $ 0.390    $ & $ 0.106    $\\%&     -      \\
   &25,54 & $ 0.110    $ & $   -      $\\%&     -      \\
   &25,55$^\ddagger$
          & $ 0.351    $ & $ 0.272    $\\%&    1.97    \\
Fe &26,54 & $ 0.894    $ & $  1.96    $\\%&    2.63    \\
   &26,55$^\dagger$
          & $ 0.514    $ & $ 0.329    $\\%&     -      \\
   &26,56 & $  9.02    $ & $  21.1    $\\%&    1.73    \\
   &26,57 & $ 0.349    $ & $ 0.784    $\\%&    2.65    \\
   &26,58$^*$ 
          & $ 2.582\times10^{-2}$ & $ 3.839\times10^{-2}$\\%&   0.845    \\
%   &26,60 & $   -      $ & $   -      $\\%&     -      \\
Co &27,56 & $ 1.374\times10^{-2}$ & $   -      $\\%&     -      \\
   &27,57$^\dagger$
          & $ 3.250\times10^{-2}$ & $   -      $\\%&     -      \\
   &27,59$^*$ 
          & $ 1.024\times10^{-2}$ & $ 1.089\times10^{-2}$\\%&   0.311    \\
Ni$^*\!$
   &28,56 & $ 4.479\times10^{-4}$ & $   -      $\\%&     -      \\
   &28,58 & $ 0.321    $ & $ 0.833    $\\%&    1.63    \\
   &28,59 & $ 1.826\times10^{-2}$ & $ 2.205\times10^{-2}$\\%&     -      \\
   &28,60 & $ 0.144    $ & $ 0.359    $\\%&    1.76    \\
   &28,61 & $ 6.345\times10^{-3}$ & $ 9.529\times10^{-3}$\\%&    1.01    \\
   &28,62 & $ 1.831\times10^{-2}$ & $ 4.411\times10^{-2}$\\%&    1.50    \\
%   &28,64 & $   -      $ & $   -      $\\%&     -      \\
\noalign{\vspace{2pt}} \hline
\end{tabular}
\end{minipage}
%\smallskip \\ %\vspace{-0.5\baselineskip}
{\footnotesize 
$^*$Source abundances error can be as large as 100\%.\\
$^\dagger$Propagated abundance over-estimated since K-capture.\\
$^\ddagger$Propagated abundance under-estimated as K-capture daughter nucleus.}
\vspace{-0.5\baselineskip}
\end{table*}%=====================================================+

Table 2 lists propagated and source isotopic abundances at 200 MeV/nucleon
($\Phi=450$ MV) for $z_h=4$ kpc. All abundances are normalized to $^{16}$O.
The Source/Solar System elemental abundance ratios are shown in
Figure 9, where the solar system abundances have been taken from 
Anders and Grevesse (1989). 
Our derived source abundance of Ti relative to Fe seems to be  high,
while the propagated abundances (Figures 1, 3) are satisfactory. This source abundance is however very sensitive to  the production cross sections.
Note also that since we use consistent propagation parameters for the entire
range of nuclei rather than adjusting for separate groups as is often done,
 this reduces our freedom to adjust the source abundances.
Source abundances of largely secondary nuclei $^{33}$S, $^{41}$Ca, $^{50}$Cr,
$^{58}$Fe, $^{59}$Co, and isotopes of Cl, Ti and Ni are small and 
also essentially depend on the production cross sections, and
the error here thus may be as large as 100\%.
The typical overall error of our source abundances for other nuclei can be estimated
as about 20\% based on conservative estimates of the cross section 
errors.

The process of electron K-capture from the ISM is not currently included so
that  K-capture isotopes such as $^{49}$V, $^{51}$Cr, $^{55}$Fe, 
and $^{57}$Co are over-estimated by 20--50\%, while their 
daughter nuclei $^{49}$Ti, $^{51}$V, $^{54}$Cr, $^{55}$Mn are under-estimated
by 20--100\%. Inclusion of the process of electron K-capture will
improve their isotopic abundances and will allow study of
the CR acceleration time scale. Since production and decay of
K-capture isotopes and radioactive isotopes depends critically on 
the matter density  distribution, our next goal is therefore
to include this process in the model.
More details will be given in a forthcoming paper.

The relative isotopic fraction
within an element for the purely secondary nucleus B, 
and secondary isotopes of Ca, Ti, V, and Co are in  good agreement
with measurements (B: Lukasiak et al.\ 1999).
The same is true for abundances of pure secondary F and Sc. As already mentioned,
$^7$Be is slightly under-predicted and $^9$Be is over-predicted
compared to measurements (Connell 1998, Lukasiak et al.\ 1999).

The current elemental abundances were tuned
to be consistent with low energy measurements by Ulyss\-es (DuVernois
and Thayer 1996). The isotopic composition of Cl
has been tuned to Ulysses data (Connell et al.\ 1998),
Ca, Ti, V, Cr, Co to Voyager data (Lukasiak et al.\ 1997a,b),
S, Mn, Ni to Ulysses (Connell and Simpson 1997, DuVernois 1997, 
Thayer 1997) and Voyager data (Lukasiak et al.\ 1997a,b, 
Webber et al.\ 1997, Lukasiak et al.\ 1999),
C, N, O, Ne, Mg to Voyager (Webber et al.\ 1996, Webber et al.\ 1997) and 
CRRES data (Duvernois et al.\ 1996), and
Si and Fe using Voyager (Lukasiak et al.\ 1997a, Webber et al.\ 1997), 
ALICE (Hesse et al.\ 1996), 
CRRES data (Duvernois et al.\ 1996), and Ulysses data (Connell and Simpson 1997). 

This is however a preliminary result in the sense that there is
still a problem with tuning elemental and isotopic abundances 
over  the whole energy range. There is some disagreement with the
elemental abundances at higher energies by HEAO-3 (Engelmann et al.\ 1990),
which is not surprising considering that the measurements
show somewhat different spectral indices for different isotopes while
we use a single power-law injection index.
On the other hand, the isotopic composition measurements are carried out
at energies which are strongly affected by solar modulation. Employing 
a model which is better than a simple force-field approximation
will be vital for further analysis.
Our neglect of K-capture  is another reason why these results are preliminary
and intended to illustrate the method rather than provide a basis for interpretation.
% intended to illustrate the method rather than provide the basis for
%interpretation..
%%% Figure 10 %%%

\begin{figure*}[!tb]
\centerline{\includegraphics[width=110mm,height=100mm]{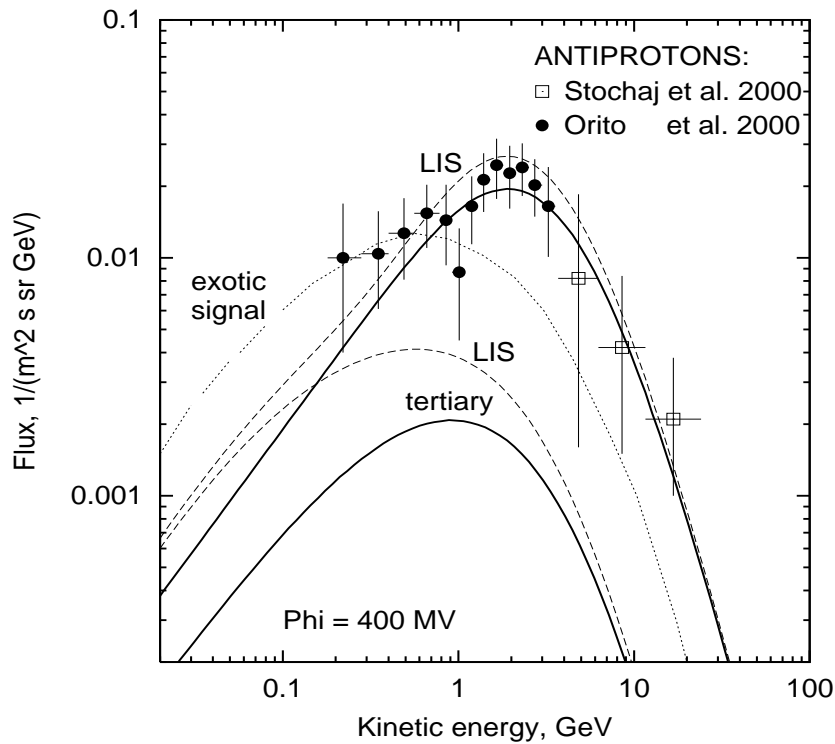}} \vspace{-1.5em}
\caption{Calculated antiproton LIS and modulated spectrum with data:
BESS -- Orito et al.\ (2000), MASS91 -- Stochaj et al.\ (2000).
The top curves are the total background. 
The ``tertiary'' components (LIS and modulated) are shown separately. 
Also shown is an example exotic signal Bergstr\"om et al.\ (1999) 
extended to lower energies.}
\end{figure*}

\section*{SECONDARY ANTIPROTONS}

Secondary positrons and antiprotons in Galactic CR and some part of the
diffuse Galactic gamma rays are produced in collisions of CR
particles with interstellar matter\footnote{ Contributions from
possible nearby source(s), black hole evaporation,
and WIMP annihilation into positrons and antiprotons are also discussed. }.
Because they are secondary, they reflect the {\it large-scale} nucleon
spectrum independent of local irregularities in the primaries and thus
provide an essential check on propagation models and also on the
interpretation of diffuse gamma-ray emission
(Strong et al.\ 2000).  These are an important diagnostic for
models of CR propagation and provide information complementary to that
provided by secondary nuclei.  However, unlike secondary nuclei,
antiprotons and positrons reflect primarily the propagation history of
the protons, the main CR component.

Secondary antiprotons have been computed in the same model
as for the nuclei. The halo size has been set to 4 kpc; the exact
value is unimportant for antiproton calculations provided the
propagation parameters are tuned to match B/C ratio.
The adopted nucleon injection spectrum, after propagation,
match\-es the local one. 
We calculate antiproton production and propagation using the basic formalism
described in Moskalenko et al.\ (1998). To this we have added antiproton annihilation 
and treated inelastically scattered antiprotons as a separate 
``tertiary'' component. 
The antiproton production by nuclei with $Z\geq2$ is calculated in two ways:
employing scaling factors as described in Moskalenko et al.\ (1998), and using effective
factors given by Simon et al.\ (1998), who make use of the DTUNUC code, 
and which appears to be more accurate than simple scaling.
(The use of the Simon et al.\ factors is consistent since their adopted
proton spectrum resembles our spectrum above the antiproton production threshold.) 
The effect on the antiproton flux at low energies is
however small, and the two approaches differ by about 15\%. 

The results are shown in Figure 10 (Moskalenko et al.\ 2001). 
The upper curves are the local interstellar flux and the lower are modulated. 
The two lowest curves show separately the contribution of 
``tertiary'' antiprotons, which is the dominant component at low energies. 
There still remains an
excess for the lowest energy points at the 1$\sigma$ level which
may indicate an additional component.

\section*{CONCLUSIONS}

It is now possible to make a spatially resolved treatment of a 
full reaction network with a realistic model of the Galaxy and CR propagation.
Reacceleration with a slighly larger index for the diffusion 
coefficient than the Kolmogorov law fits B/C, sub-Fe/Fe and CR antiprotons
quite well over the whole energy range. The halo size
indicated by the four radioactive isotope ratio measured by ACE is in the
range 3 -- 7 kpc, but the dispersion between the isotopes is large presumably 
due to cross-section inaccuracies. In fact 
the CR measurements have now become relatively more accurate than the cross section 
measurements, so that the latter become a factor restricting further progress.

To remove uncertainty connected with elemental abundances errors 
and denominator production cross sections, we propose
in future to use ratios like radioactive secondary/major primary, namely \linebreak[4]
$^{26}$Al/$^{28}$Si, $^{36}$Cl/$^{56}$Fe, and $^{54}$Mn/$^{56}$Fe instead
of widely used \linebreak[3] $^{26}$Al/$^{27}$Al, $^{36}$Cl/Cl, and $^{54}$Mn/Mn.
The ratio  \linebreak[3] $^{54}$Mn/$^{56}$Fe is probably the best candidate because there 
are only very minor contributions from other nuclei.
In the case of Be, the ratio $^{10}$Be/$^{12}$C is probably  better than
$^{10}$Be/$^9$Be.  Finally,
the accurate measurement of all isotopic abundances by one experiment, e.g. ACE, 
will help greatly to constrain the problem. 

\section*{ACKNOWLEDGEMENTS}

The authors thank F.\ Jones, V.\ Ptuskin, and A.\ Lukasiak for helpful
discussions, and S.A.\ Stephens and R.A.\ Streitmatter for
providing the database of B/C measurements. 
I.\ Moskalenko acknowledges support from NAS/NRC Senior Research 
Associateship Program.

\end{document}